# Fast and sensitive detection of hemoglobin and other paramagnetic species using coupled charge and spin dynamics in strongly fluorescent nanodiamonds

F. Gorrini[1]*, R. Giri[1], C. E. Avalos[2], S. Tambalo[1], S. Mannucci[3], L. Basso[1,4], N. Bazzanella[4], C. Dorigoni[1], M. Cazzanelli[1,4], P. Marzola[5], A. Miotello[4] and A. Bifone[1]

[1] Center for Neuroscience and Cognitive Systems, Istituto Italiano di Tecnologia, Corso Bettini 31, 38068 Rovereto, Trento, Italy

[2] Institut des sciences et ingénierie chimiques, Ecole Polytechnique Fédérale de Lausanne, 1015 Lausanne, Switzerland

[3] Department of Neuroscience, Biomedicine and Movement Sciences, University of Verona, Strada Le Grazie 8, 37134, Verona, Italy

[4] Department of Physics, University of Trento, 38123, via Sommarive 14, Povo, Trento, Italy

[5] Department of Computer Science, University of Verona, Strada Le Grazie 15, 37134 Verona, Italy

## Abstract

Sensing of a few unpaired electron spins, such as metal ions and radicals, is a useful but difficult task in nanoscale physics, biology, and chemistry. Single nitrogen-vacancy (NV) centers in diamond offer high sensitivity and spatial resolution in the optical detection of weak magnetic fields produced by a spin bath, but often require long acquisition times, of the order of seconds. Here we use a novel approach, based on coupled spin and charge dynamics in dense NV ensembles in strongly fluorescent nanodiamonds (NDs) to sense external magnetic dipoles. We apply this approach to various paramagnetic species, including gadolinium complexes, magnetite nanoparticles and hemoglobin in whole blood. Taking advantage of the high NV density, we demonstrate detection of subnanomolar amounts of Gd spins with ultrashort acquisition time approaching 10 ms. Strong luminescence, high sensitivity and short acquisition time make dense NV$^-$ ensembles in NDs a powerful tool for biosensing and bioimaging applications.

# Introduction

Negatively charged nitrogen-vacancy (NV⁻) centers in diamond have been extensively used to measure magnetic fields[1,2], electric fields[3,4] and as temperature sensors[5,6] at the nanoscale. Owing to their biocompatibility and ease of functionalization, fluorescent nanodiamonds (NDs) are promising as optical imaging probes for biological applications[7], such as in-vivo tracking[8,9], magnetic imaging[10] and nanothermometry[11]. More recently, diamond-based nanosensors have been proposed for detection of magnetic dipole interactions by measuring the effects of fluctuating magnetic fields generated by electron spins on the longitudinal spin relaxation time ($T_1$) of NV centers[12]. Potential applications of $T_1$ relaxometry with NV centers in NDs include structure determination of biomolecules labeled with paramagnetic ions, monitoring the accumulation of paramagnetic species in tissues and bodily fluids, and detection of reactive oxygen species at very low concentration in living cells as a potential marker of cellular viability, stress or aging. Unlike coherent detection methods normally applied for magnetometry, relaxometric measurements can be performed with all-optical detection schemes without the need to manipulate the NV⁻ spin states with microwaves, which are strongly absorbed by biological samples.

Detection of paramagnetic species using single or dilute ensembles of NVs in diamond has been demonstrated with gadolinium[13-17], manganese[18] and ferritin proteins[18-20]. Denser ensembles of NV centers may be desirable to increase signal-to-noise (SNR) ratio in photon detection and to reduce measurement time while retaining high spatial resolution, thus making these promising techniques viable for routine application in biomedicine. However, high concentrations of NVs, defects and substitutional impurities may result in drastically shorter coherence times and broader spectral lines, and also favor charge state conversion of NV centers[21,22].

Recently, longitudinal spin relaxation was investigated in strongly fluorescent bulk diamond with high concentration of nitrogen defects (200 ppm) and NV centers (10 ppm)[23]. The evolution of the spin-polarized state in this regime showed a complex behavior, reflecting the contributions of two different mechanisms involving spin and charge dynamics. Indeed, the preparation pulse can induce polarization of the spin state of NV⁻ centers as well as charge state conversion from NV⁻ to NV⁰ [21,24-29]. The subsequent evolution depends on the interplay between spin relaxation and recharging in the dark, with a paradoxical inversion of the exponential profile when charge conversion dominates the dynamics. Whether effective sensing of magnetic interactions can be performed in a regime of intense charge conversion, however, remains unclear.

Here we study the effect of fluctuating magnetic noise in the gigahertz range, induced by paramagnetic agents, on spin and charge dynamics in fluorescent NDs with dense NV ensembles. We exploit these complex dynamics for sensing magnetic dipolar interaction with (a) gadoteridol, a paramagnetic chelate complex of $Gd^{3+}$ widely used as contrast agent for diagnostic MRI; (b) magnetosomes (MNs)[30,31], superparamagnetic nanostructures of magnetite naturally synthetized by magnetotactic bacteria; (c) blood, containing paramagnetic deoxygenated hemoglobin. We demonstrate detection of subnanomolar amounts of paramagnets, with ultrashort acquisition time approaching 10 ms, a benefit of the high concentration of NV centers. The combination of strong fluorescence, excellent sensitivity to magnetic interactions and purely optical irradiation makes highly fluorescent NDs ideal candidates for applications in biological systems.

## Results

We used three types of highly fluorescent NDs (Bikanta-Berkeley, CA): unfunctionalized NDs with approximate diameter of 100 (FND100) and 40 nm (FND40), and 100 nm NDs coated with a ≈10 nm layer of silica (SiFND100). All samples have high concentration of NV centers (about 5 ppm) with short coherence time in the range 50-80 ns (see Table 1 in Supplementary Material section). In Fig. 1a we present the pulse sequence used to measure the $T_1$ relaxation dynamics. The ground state of the NV center is a spin triplet, with the $m_s = \pm 1$ levels upshifted by 2.87 GHz (the zero-field splitting, ZFS) with respect to the $m_s = 0$ level. The first 532 nm laser pulse (1 ms, 2 mW) populates the $m_s = 0$ level preferentially through a spin dependent transition from the excited state through the metastable singlet states[32]. As a consequence, the fluorescence (FL) of the $m_s = \pm 1$ levels is lower than that of the $m_s = 0$ level. In the variable dark time $\tau$ the system relaxes towards the equilibrium condition. The characteristic timescale for this mechanism is known as the longitudinal spin relaxation time ($T_1$). The mechanisms that influence $T_1$ are interactions with phonons[33], NV-NV dipolar coupling[34,35] and surface charges[15]. In addition to these mechanisms $T_1$ can also be reduced by magnetic noise resonant with the $|0\rangle \leftrightarrow |\pm 1\rangle$ transitions[12], such as the one produced by a spin bath. The second laser pulse of much shorter duration (1 μs) was used to read-out the remaining polarization. A reset time of about 5 ms was applied to allow the system to reach charge equilibrium before another pumping pulse was applied. No microwaves were applied in this sequence.

In Fig. 1b we show the exponential decay profile of FL from the unfunctionalized 100 nm NDs. We measured $T_1 \approx 0.41$ ms, slightly shorter than in bulk diamond samples with similar concentrations of NVs[36], likely a result of the nanoscopic size[15]. When NDs are mixed with gadoteridol (see Materials and Methods) the NVs interact with the Gd spins surrounding the NDs through a magnetic dipole-dipole interaction. The effects of these interactions with gadoteridol at various amounts are shown in Fig. 1c and 1d. With increasing amount of Gd, a second component appears and becomes more prominent, while the exponentially decaying component becomes steeper. At the highest amounts of Gd, a paradoxical inversion of the curve was observed, with FL building up with increasing dark time (Fig. 1d). This Gd-induced behavior can be qualitatively described in terms of dynamics of spin and charge at different timescales[23,29]. The exponentially decaying component represents the $T_1$ relaxation of the spin-polarized $m_s = 0$ level. However, the preparation pulse can also affect the charge state of the NV⁻ centers. The recovery of FL on longer timescales can be attributed to recharge of NV⁰ centers in the dark, as was observed in the case of bulk diamond[23]. As $T_1$ decreases because of increasing magnetic interactions, recharging in the dark gradually dominates the evolution of the FL signal.

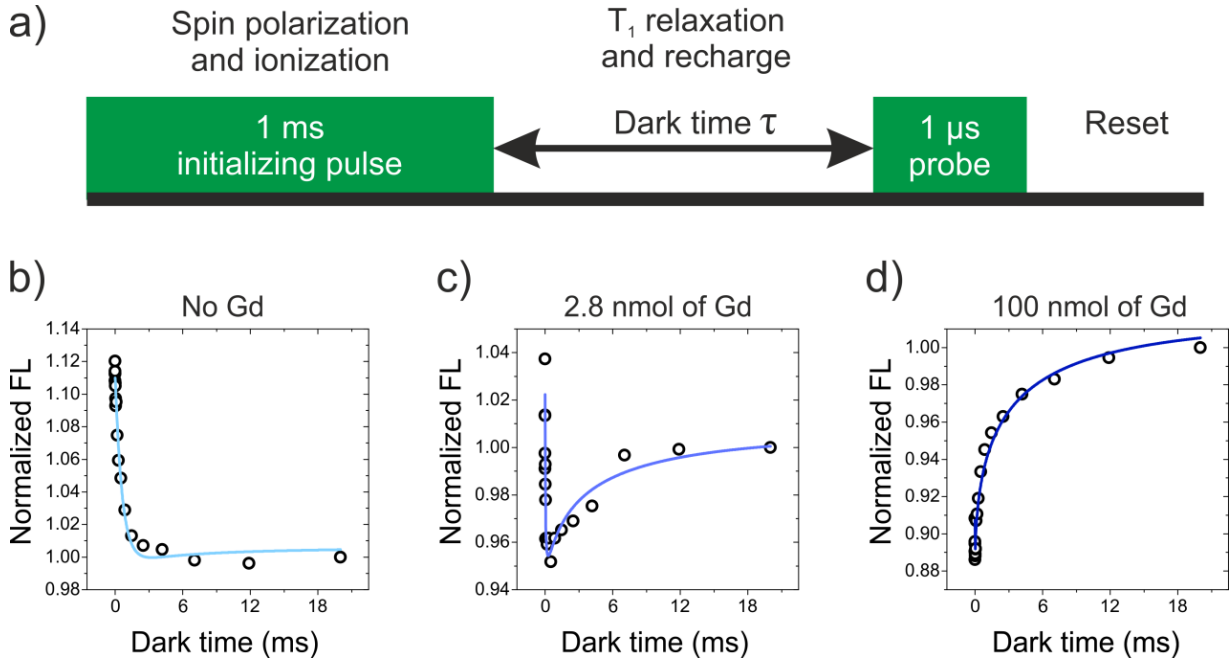

**Figure 1 | Time evolution of FL as a function of dark time and amount of gadoteridol.** (a) The pulse sequence consists of an initializing pulse, a variable dark time τ, and a probing pulse. The initializing pulse polarizes part of the NV centers and ionizes other NV centers. Then, during the dark time, the spin system relaxes and recharge takes place. The probing pulse detects the level of FL after a time τ. Evolution of FL in the dark from samples containing $\approx 2 \times 10^9$ NDs without Gd, mixed with 2.8 nmol of Gd and mixed with 100 nmol of Gd are reported in (b), (c) and (d), respectively. The data were fitted with the function described by equation (1) (blue curves). The fluorescence decay and rise are attributed to spin and charge dynamics, respectively. For higher amounts of Gd, a complete inversion of the FL evolution curve is observed, as the dynamics become dominated by recharging in the dark.

To estimate the parameters pertaining to the two competing dynamics, we empirically fit the FL profile with a sum of two stretched exponentials[23]:

$$I(t) = I_{eq}\left[1 - \alpha e^{-\left(\frac{t}{T_r}\right)^m} + \beta e^{-\left(\frac{t}{T_1}\right)^n}\right] \quad (1)$$

where $I_{eq}$ is the equilibrium value of FL at long times (≈20 ms), $T_r$ is the recharge time, $m$ and $n$ are two stretching exponents and $\alpha$ and $\beta$ are two pre-exponential coefficients. The stretching exponents $n$ and $m$ account for the distribution of relaxation times $T_1$ and of recharge times $T_r$, respectively, in the ensemble of NV centers. Therefore, all the parameters of equation (1) are mean values, and $T_1$, in particular, is an effective value of the longitudinal spin relaxation time (a full derivation can be found in Supplementary Material). Fit curves described by equation (1) are shown in blue color in Fig. 1b-d.

The longitudinal relaxation time $T_1$ was extracted from the fits and plotted in Fig. 2a, as a function of $n_{Gd}$, the amount of gadoteridol mixed with NDs. Optimal fit was obtained for $n$ in the range $0.5 - 1.0$. $T_1$ decreased by two orders of magnitude with increasing amounts of Gd, from 410 µs (red circle on the y-axis) to <5 µs. Shortening of T1 can be attributed to high-frequency magnetic noise[13-17] produced by the fluctuating spin bath of Gd ions that induces transitions between the states $|0\rangle$ and $|\pm 1\rangle$. If the Gd-Gd interaction is of dipolar type, the relation between $T_1$ and the baths parameters is

$$\frac{1}{T_1} = \frac{1}{T_1^0} + \frac{3\gamma_e^2}{2\pi}\frac{\langle B_\perp^2\rangle f_{Gd}}{f_{Gd}^2 + D^2} \quad (2)$$

where $D$ is the ZFS, $T_1^0 = 410$ µs is the longitudinal relaxation time of the NV centers without Gd, $\langle B_\perp^2 \rangle$ is the transverse magnetic field variance (with zero mean) produced by the paramagnetic environment with frequency $f_{Gd}$, and $\gamma_e \approx 2$ is the electron gyromagnetic ratio. Theoretical predictions based on equation (2) are in good agreement with experimental data (blue curve in Fig. 2a). Purely magnetic noise is unlikely to affect charge dynamics, and the related parameters $T_r$ and $m$ should not vary. We determined $T_r = 3$ ms with $m = 0.5$ from the fit of experimental data (see the Discussion section). The pre-exponential coefficients $\alpha$ and $\beta$ depend on the laser pulse parameters as well as on the relaxation mechanisms and were also determined from the fit. The difference $\beta - \alpha$ (Fig. 2b) corresponds to the FL contrast at t=0, immediately after the initializing pulse (equation (1)). The coefficient $\alpha$ is related to the ionization process and increases slightly with $n_{Gd}$, while $\beta$ depends mainly on the degree of polarization and decreases with $n_{Gd}$ ($\alpha$ and $\beta$ not shown individually here). Therefore, low values of $\beta - \alpha$ at high $n_{Gd}$ are indicative of low polarization levels. Indeed, a consequence of short $T_1$ and small $\beta$ values is fast depolarization of the $m_s = 0$ level, thus reducing the net polarization achieved by the initializing pulse for a certain laser power and pulse duration (see Supplementary Material). The reduced polarization of $m_s = 0$ in the presence of Gd is reflected in the reduction of the optically detected magnetic resonance (ODMR) contrast (Fig. 2c, see also Materials and Methods). The FL profile reaches a minimum, $t_{min}$, when the spin-related decreasing component intercepts the charge-related increasing component. In Fig. 2b we show the variation of $t_{min}$ with respect to $n_{Gd}$. One can note the log scale for $t_{min}$, a parameter that varies by over two orders of magnitude in the range of Gd amounts explored here, providing a very sensitive measure of the number of paramagnets.

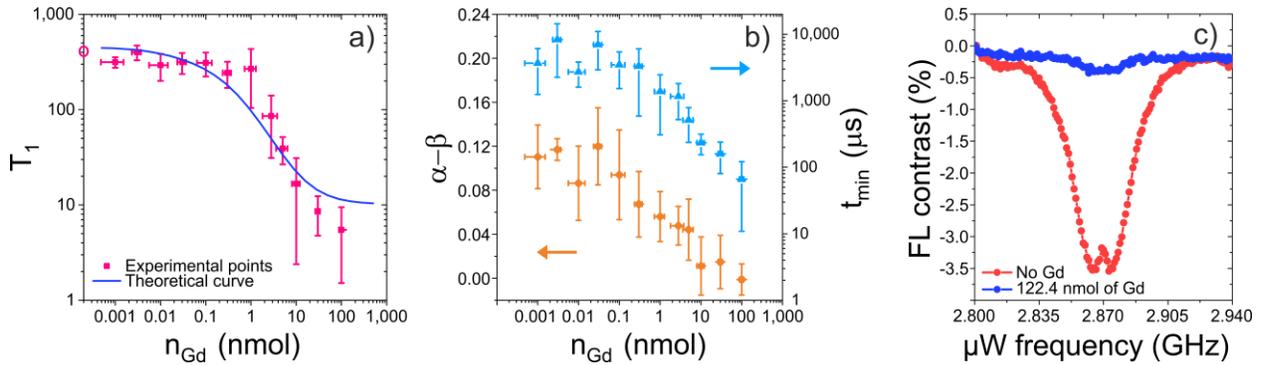

**Figure 2 | Parameters extracted from fit of FL decay curves.** (a) Longitudinal spin relaxation time (pink points) is plotted as a function of the amount of deposited Gd. The empty circle on the y-axis indicates the baseline $T_1$ of NV centers in NDs without Gd. The characteristic time for recharge is 3 ms, assumed to be constant throughout the range of $n_{Gd}$ considered here. Relaxing this assumption does not qualitatively change the results (see Discussion section). The blue curve represents the theoretical dependence of $T_1$ on Gd amount calculated from equation (2). (b) Difference between the pre-exponential coefficients $\beta - \alpha$, in orange, and position of the minimum of FL, in blue, as a function of amount of Gd deposited. Reported values in (a) and (b) are averaged over 10-20 acquisition and standard deviation sets the error bars. (c) ODMR spectra of NDs without Gd, and in the presence of > 100 nmol of Gd. With Gd, FL contrast is drastically reduced, indicative of low level of polarization obtained by the preparation pulse, and a consequence of the fast $T_1$ relaxation induced by paramagnetic interactions.

Analogous experiments were performed with 40 nm uncoated NDs and 100 nm silica-coated NDs to explore potential size and surface effects. Similarly to the uncoated 100 nm ND, we observed reduced ODMR contrast and shortening of $T_1$ with increasing amounts of Gd (Fig. 3a-f). Also in this case, large amounts of Gd resulted in an inversion of the FL profile (Fig. 3b and 3d). Additionally to the known effect of magnetic noise on spin relaxation, electric noise at the surface of ND, e.g. arising from metal ions like $Gd^{3+}$, could in principle affect the charge dynamics of the NV centers and alter the evolution of FL. For gadoteridol, however, the $Gd^{3+}$ charge is compensated by the charges carried by the chelating agent, and the complex is overall neutral. Moreover, the same behavior of the FL signal appears with the silica coated NDs, where the silica layer completely shields the diamond from the external charges and the minimum distance between the Gd chelate and the NV centers is 10 nm. This supports the idea that long-distance magnetic interactions dominate the FL signal evolution, and that electric noise is unlikely to contribute significantly.

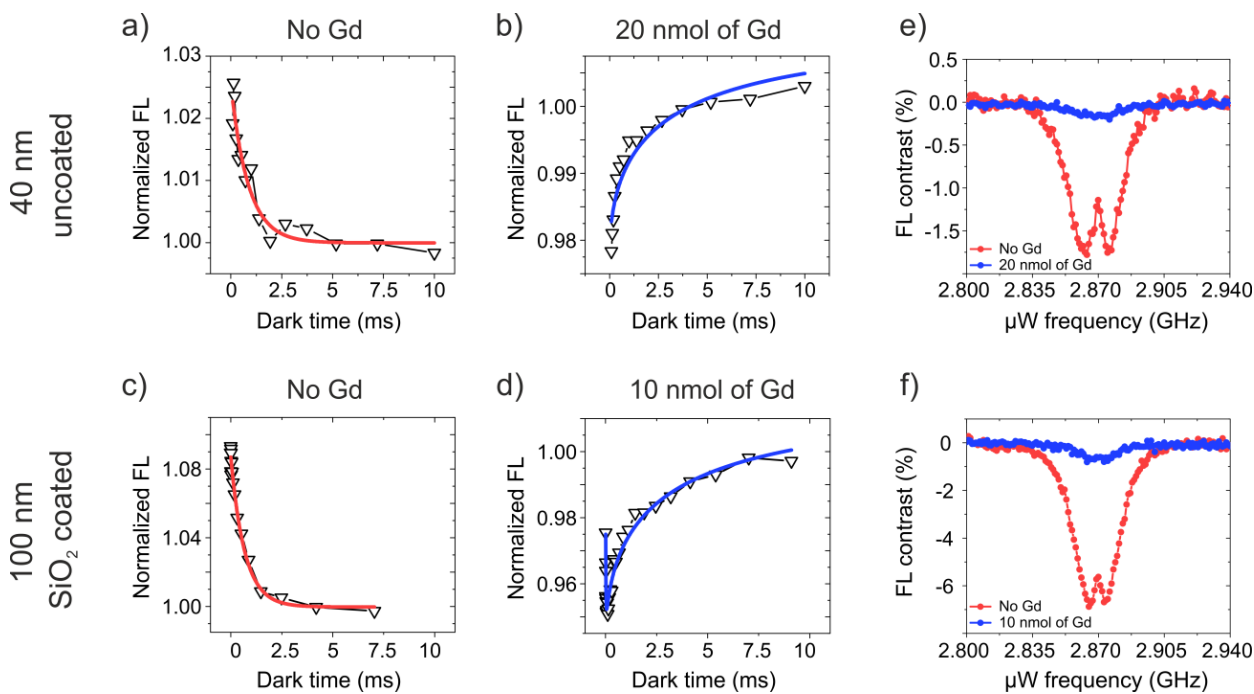

**Figure 3 | Time evolution of FL for different NDs size and coating, and ODMR.** (a,b) Spin-charge relaxation dynamics, with and without Gd in the 40 nm unfunctionalized NDs, and (c,d) in the 100 nm SiO$_2$-coated NDs. Curves are similar to the ones obtained with 100 nm uncoated NDs and were fit with a double stretched exponential. For these NDs also complete reversal of the FL curve was observed at the highest Gd amounts hereby explored. (e,f) The ODMR contrast indicates rapid depolarization of the NV centers in the presence of Gd.

We also tested the effects of magnetosomes (MNs) as a different magnetic species. MNs are nanoparticles of magnetite ($Fe_3O_4$) with an average size of 40 nm (estimated from transmission electron microscopy, TEM) and a single magnetic domain[30,31] that lends superparamagnetic properties. Fig. 4a is a scanning electron microscope (SEM) representative picture of 100 nm unfunctionalized NDs (in green) mixed with MNs (in orange) on a silicon substrate. The average magnetic field experienced by the NVs is zero, as the ODMR spectrum shows no broadening or shift, but only a reduction in contrast (Fig. 4b). With increasing amounts of MNs, the same qualitative behavior observed with Gd emerges (Fig. 4c), with faster spin relaxation and inversion of the FL profile. This effect can be explained considering that the magnetic moment of each magnetosome inverts direction over a timescale of $\tau_N$, known as the Néel relaxation time, as a result of thermal fluctuations[37]. The magnetic field produced by a single nanoparticle fluctuates at a rate

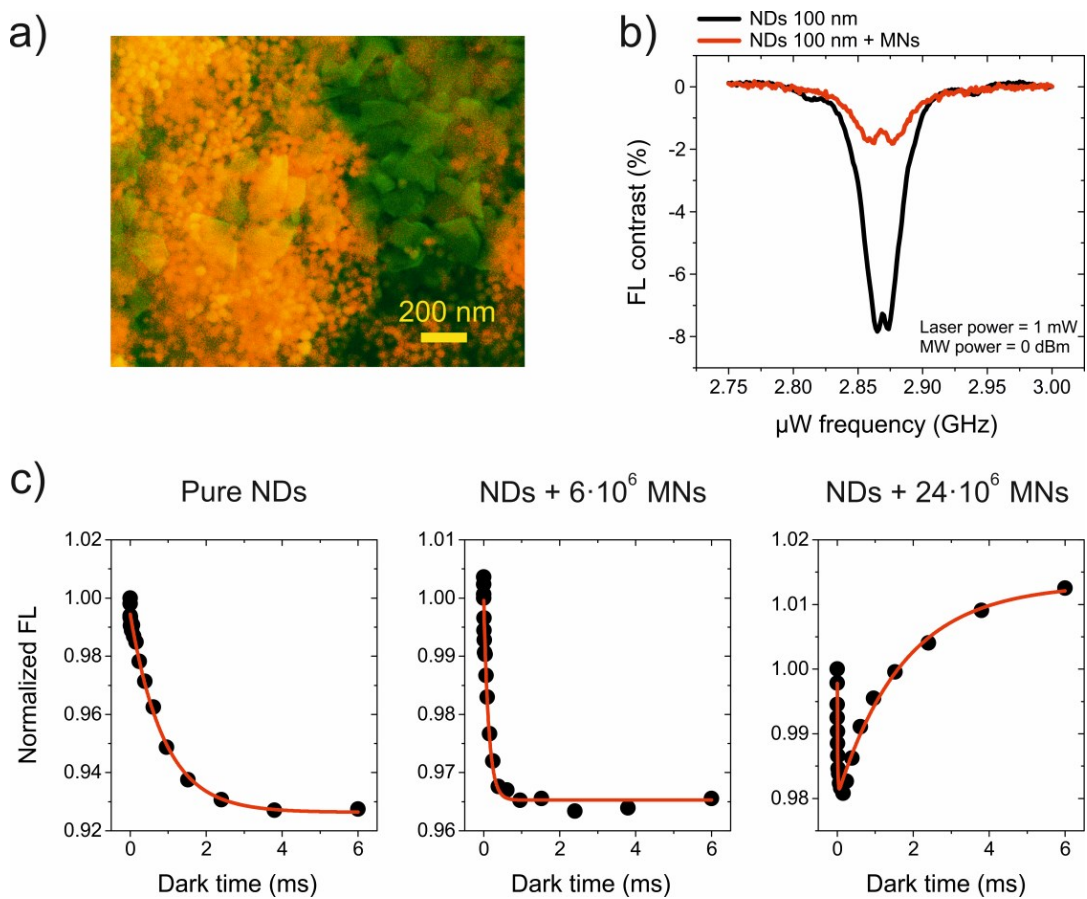

**Figure 4 | Time evolution of FL with MNs.** (a) SEM image of NDs mixed with MNs. Backscattered and secondary electrons are used to distinguish NDs (in green) from MNs (in orange). (b) Interactions with MNs induce reduction of the ODMR and (c) shortening of $T_1$ with a gradual reversal of the FL curve, similar to Gd. These findings corroborate the idea that the anomalous decay profile of the NV⁻ FL originates from magnetic interactions with the paramagnetic agents.

$$\frac{1}{\tau_N} = \nu_0 \exp\left(-\frac{KV}{kT}\right) \qquad (3)$$

where $\nu_0 \approx 1 \div 10$ GHz is the *attempt frequency*[38], $V$ is the volume of the MN and $K = 1.1 \times 10^4$ Jm$^{-3}$ the anisotropy energy constant for magnetite[39]. Given the exponential dependence on the volume of equation (3), the smaller MNs produce a fluctuating magnetic field in the GHz region, with a component at the ZFS. In line with previous results, the recharge dynamics dominates at high amounts of the superparamagnetic agent, when $T_1$ is considerably shorter.

Finally, we tested the sensing capabilities of NV ensembles in the presence of blood, a weak paramagnetic system. Red blood cells are packed with hemoglobin, and the heme groups in the deoxygenated state are paramagnetic, with a magnetic moment of $\approx 5.4$ Bohr magnetons per each heme group[40,41]. Freshly drawn rat blood was mixed with heparin and diluted with either a saline solution or distilled water (see Materials and Methods). Then 100 nm unfunctionalized NDs were added to these solutions before depositing on a glass slide. While isotonic saline solution preserves the integrity of the erythrocytes, deionized water induces hemolysis and the release of hemoglobin. Hence, we could assess the effects of hemoglobin segregated in the red blood cells or dispersed in the plasma, where it could interact directly with the ND surface. The blood/water curve of Fig. 5c decays more rapidly than the blood/saline curve (Fig. 5b), suggesting a high exposure of NDs to deoxyhemoglobin, as expected. Both curves decay faster than the reference curve acquired without blood (Fig. 5a). In both cases, the reduction of $T_1$ was detectable but insufficient to expose the effects of recharging in the dark. Several reasons may explain the lack of inversion of the fluorescence curve in the presence of blood compared to the other magnetic agents investigated in this study. First, the magnetic moment of Fe$^{2+}$ ions is $\approx 5.4\,\mu_B$, slightly less than $\approx 7.9\,\mu_B$, the value of Gd$^{3+}$ magnetic moment. Second, hemoglobin is a large molecule (about 5 nm[42]). By comparison, gadoteridol has a molecule size of $\approx 1.1$ nm. The magnetic noise increases with the magnetic moment and decreases with distance, so Fe$^{2+}$ ions are farther and cannot produce the same noise amplitude on the NVs. Third, deoxygenated, paramagnetic hemoglobin is a small fraction of the total hemoglobin, which is prevalently in the oxygenated diamagnetic form[43]. Sample preparation and optical experiments were performed in air, and hemoglobin is likely to be predominantly in the oxygenated state (>96%).

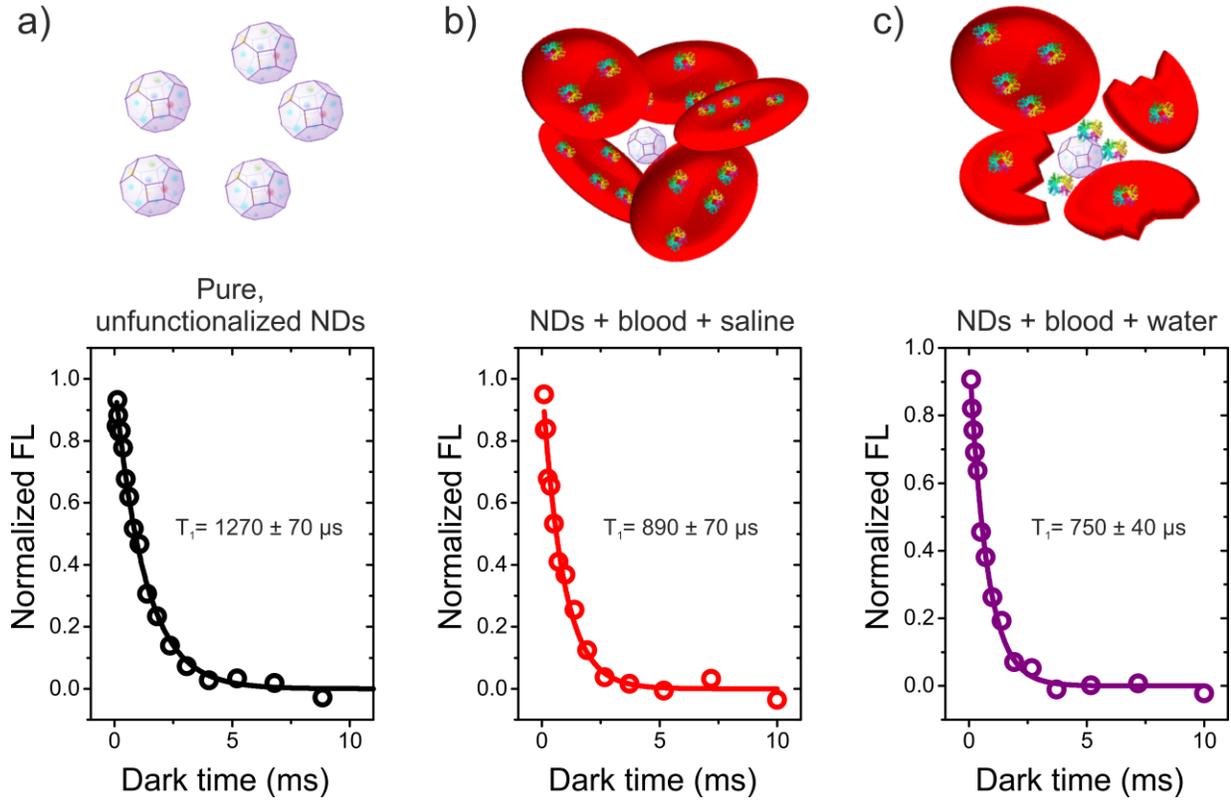

**Figure 5 | FL decay profile of unfunctionalized NDs mixed with blood.** (a) FL decay of pure NDs, (b) NDs mixed with blood/saline solution and (c) NDs mixed with blood/distilled water. Water is non-isotonic and induces hemolysis and release of paramagnetic deoxyhemoglobin. This leads to further reduction of $T_1$, compared to the blood/saline case. No inversion of the curve was observed under these conditions, in which the evolution of the curve is dominated by spin relaxation.

## Discussion

Photo-induced charge conversion from $NV^-$ to $NV^0$ [21] and from $NV^0$ to $NV^-$ [44] has been reported. Conversion of the NV charge state was shown to depend on laser wavelength[25], on the distance from the surface[28], and on the presence of other defects[22]. In our NDs, most of the NVs lie several nanometers from the surface and are in the negative charge state. Therefore, the process is likely to be dominated by electron transfer from $NV^-$ to neighbor charge trapping defects via laser-induced ionization, and from the defects to the $NV^0$ via tunneling in the dark, with no need of optical or thermal excitation. This tunneling-mediated recharge has been recently observed in bulk diamond[23,29,35] and is responsible for the $NV^-$ replenishment, explaining the rising component of FL during the dark time (Fig. 1c and 1d). The two alternative paths of ionization/recharge and polarization/relaxation are indicated in the scheme of Fig. 6a.

With an appropriate combination of bandpass filters, we recorded the FL of NV⁻ ($750 - 800$ nm range) and NV⁰ ($550 - 600$ nm range), at the same point on the sample and under the same laser initialization conditions (Fig. 6b). The increase in the NV⁻ FL, due to recharge, is mirrored by a reduction in the NV⁰ FL, with comparable timescales of $\approx 2 - 3$ ms. Importantly, the NV⁰ FL has the same decay rate with 1 nmol and 10 nmol of deposited Gd, differently from the NV⁻ FL. At the same time, the features of the NV⁻ rising component (both the timescale and exponential coefficient), seem to be independent of the type of magnetic environment (paramagnetic ions or superparamagnetic nanoparticles), provided that the noise is sufficiently strong to reduce $T_1$ severely. These observations suggest that recharge is always present, but it can be revealed only when $T_1$ decreases by about one order of magnitude. As anticipated, of all the parameters related to recharge only $\alpha$ was left unrestrained during fitting, since it depends on the initialization phase, i.e., on the competing processes of ionization and polarization, and also on the relaxation time $T_1$ (see Supplementary Material). On the contrary, $T_r = 3040$ μs and $m = 0.5$ were determined from the decay of the NV⁰ FL and fixed when fitting the NV⁻ experimental data. Conceivably, the timescale and stretched exponential shape of the rising component are peculiar to each sample, depending on the number of defects and on their distance from the NVs[21,27,29]. In our samples, the recharge time of $\approx 3$ ms is slightly longer than previously reported[23,35]. It's important to remark that $T_1$ always decreases with $n_{Gd}$, irrespectively of the detailed structure of the fitting function. In fact, we fitted the NV⁻ FL with two simple exponentials, where the parameters of recharge, as well as those of spin dynamics, were left free and did not see any qualitative difference with the behavior shown in Fig. 2a.

Because of these spin and charge dynamics, other parameters, in addition to $T_1$, can be used for a quantitative estimate of paramagnet concentration. In Fig. 2a and 2b, we showed that $T_1$, $t_{min}$ and $\beta - \alpha$ all decrease over the range of $n_{Gd}$ we explored. For the three parameters, the steepest decrease occurs between 0.1 and 10 nmol, with a tendency to saturate below 10 pmol. Taking 10 pmol as the minimum detectable amount of gatoderidol, we estimate a maximum of 550 spins of Gd³⁺ decorating each ND, or equivalently a density of 0.018 spins/nm² over the surface of each ND, a number comparable with previous reports[13,15,18]. However, it is unlikely that all Gd was at the surface of NDs, and this estimate of sensitivity is very conservative. Turning to blood, if we assume that paramagnetic deoxyhemoglobin and methemoglobin are 4% of total hemoglobin, a total of 260 spins at the most can be found in the proximity of each ND. Even if this low number of spins (together with the lower Fe²⁺ magnetic moment, compared to Gd³⁺) is not able to reveal the recharge dynamics, it can still reduce $T_1$ by a detectable amount (Fig. 5a-c).

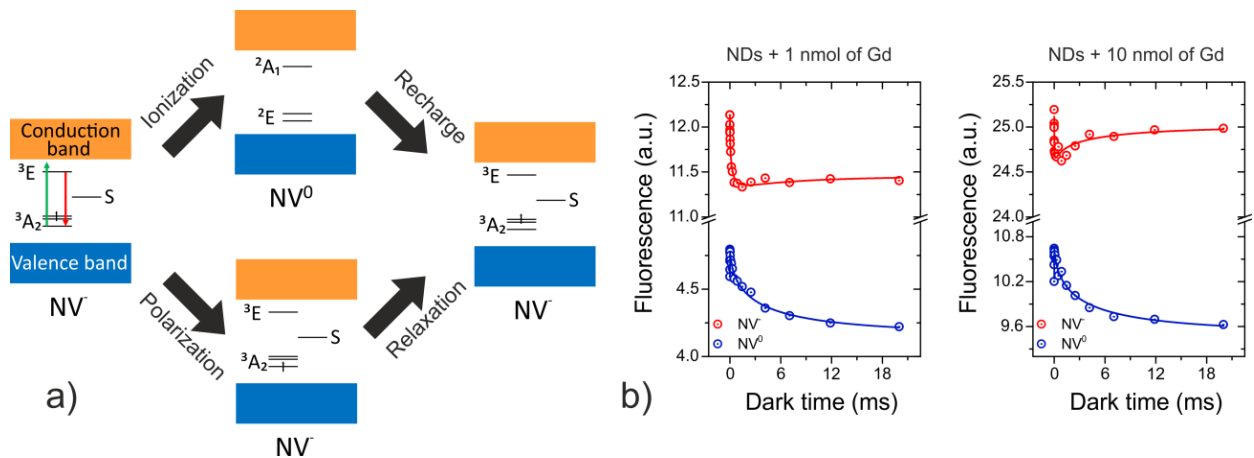

**Figure 6 | Schematics of spin and charge dynamics.** (a) After the initial pulse, an NV center can be polarized or ionized. Trap states momentarily capture photoexcited electrons. During the dark time, the system evolves toward an equilibrium configuration, while spin relaxation and recharge take place. The probing pulse detects the FL level, which depends on both the remaining polarization and the absolute number of NV$^-$ after a variable dark time. (b) Selective measurement of FL from NV$^-$ and NV$^0$ centers shows the charge-state conversion due to recharging in the dark.

A key advantage of using dense ensemble of NDs for sensing applications is fast data acquisition. The typical acquisition time needed to plot a full curve is often less than 1 min, in some cases approaching a few seconds (see Materials and Methods). Ultrafast detection schemes[13,14] rely on the so-called single-point detection, where the number of paramagnets is extracted by measuring the fluorescence drop of a single reference point. The integration time also depends on $n_{Gd}$ and is minimum when the detection point is at $\approx T_1$, the optimal value to reduce the photon shot noise (Supplementary Material). As an example of our integration time, a variation of 50% in 1 nmol of gadoteridol (corresponding to an increment of $\approx 25000$ spins for each ND) can be detected using an initialization time $\tau_{in}$ of 200 µs, a readout time $\tau_{ro}$ of 50 µs, with 35 iterations and a total acquisition time of ≈12 ms, a much shorter time compared to previous reports[13-15,17,18]. A variation of 50% in 10 pmol of gadoteridol (increment of 275 spins for each ND) can be detected in 0.5 s, which again is shorter by a factor $2-10$ than previous reports. Quantitative detection using such a short acquisition time at higher concentrations is unprecedented for this kind of detection schemes and could enable real-time tracking of changes of the paramagnetic environment in biological tissues. We note that a short acquisition time is largely favored by a high number of detected photons from dense NVs samples and by long $\tau_{ro}$. Since the readout time for NV ensembles can be much longer compared to the one of single NVs (hundreds of microseconds in the first case[23] and $\approx 300$ ns in the second case[32]) the above values lead to fast acquisition of the signal, without considerably repolarizing the NV ensemble. Fast acquisition might serve in the future to image the oxygenation level of tissues, or to track the flow of radicals and other paramagnetic species in living cells with fluorescence microscopy.

Finally, we must point out that, in the interpretation of our results, we considered only the effects of external agents, without explicitly accounting for surface charges and NV-NV cross-relaxation. Surface charges are mainly due to dangling bonds with unpaired electron spins on the diamond surface. These spins provide additional magnetic noise on the NVs and a shortening of $T_1$ which is inversely proportional to the NDs size[15]. Because of a large number of NDs within the laser spot size, this contribution is always averaged over NDs of many different sizes, independently of $n_{Gd}$. Moreover, we observed a similar $T_1$ for the three types of NDs without Gd, irrespectively of the size and surface coating, suggesting that surface charges play a minor role. In fact, surface charges might provide a substantial contribution to $T_1$ relaxation for NDs smaller than ≈20 nm[15]. Cross-relaxation is another source of $T_1$ relaxation and involves the fast depolarization of coupled NVs in a dense ensemble[35,36]. This mechanism is quenched by a static magnetic field that can remove degeneracy between differently oriented NV centers[23,34], but is hardly sensitive to the high-frequency magnetic field of a spin bath. For these reasons, any contribution to $T_1$ from surface charges and cross relaxation is not expected to depend on $n_{Gd}$.

In conclusion, we investigated the spin and charge dynamics of dense ensembles of NV centers in NDs in the presence of GHz magnetic fluctuations due to external paramagnetic agents. Our results show that the fluorescence evolution consists of two components, related to different mechanisms. The fast component, decaying with a characteristic time $T_1$, comes from spin depolarization dynamics and is strongly influenced by the amount of magnetic noise induced by paramagnetic molecules (gadoteridol and deoxyhemoglobin) or superparamagnetic nanoparticles (magnetosomes). The slow increasing component characterized by $T_r$ is due to recharging in the dark. The latter mechanism is thought to be due to tunneling-mediated charge conversion[23,29,35] from $NV^0$ to $NV^-$ during the dark time and it might be accentuated by a large number of lattice defects in these samples. Interestingly, the mechanism of recharge dominates when $T_1$ is drastically reduced. We exploited this double dynamic of spin and charge in the detection of tens of picomoles of gadoteridol, corresponding to at most 550 Gd spins per ND. Further, we showed that high fluorescence and long readout times can speed up the acquisition time down to 10 ms at subnanomolar amounts. Additionally, the minimum of the fluorescence profile and the difference of pre-exponential coefficients were found to be novel indicators of the number of paramagnets interacting with the NDs. Our results are important for understanding the fundamental physics of spin and charge dynamics of dense NV ensemble in the presence of paramagnets and for application purposes. Strongly fluorescent NDs with short acquisition time, high magnetic field sensitivity, and long-range detection ability might find future applications in quantitative bio-sensing of magnetic dipolar interactions.

## Materials and Methods

**Experimental setup**

We used a home-built microscope system to study the dynamics of depolarization of NV ensembles. An objective with a numerical aperture of 0.25 (Plan N, Olympus) focused the excitation laser (532 nm, of Coherent Verdi) to a focal area of $\approx 2.6$ μm². Power was kept in the range of $1-10$ mW. An acousto-optic modulator (MT200-A0,5-VIS, AA Opto Electronic) produced the desired pulsed laser sequence (Fig. 1a). The same objective was then used to collect the fluorescence signal, which was filtered and attenuated by a series of band-pass filters before reaching the detector (a single photon counter, Excelitas SPCM-AQRH-14- FC). Microwaves produced by a Keysight N5171B generator were amplified and selectively pulsed with a microwave switch (Mini-Circuits, ZASWA-2-50DR+). The sample sat on a custom-made copper loop, connected to the μW line. All the optical and μW pulse sequences were remotely controlled by a programmable TTL pulse generator (PulseBlaster ESR-PRO, SpinCore Technologies). An acquisition card (PCIe-6323, National Instruments) collected the experimental data.

**Optically Detected Magnetic Resonance experiments**

Continuous wave ODMR (CW-ODMR) is a technique used to determine the sublevel structure of the ground state. We continuously irradiated the NDs samples with the same 532 nm laser and with the same power (≈2 mW) as for the $T_1$ experiments, to polarize the $m_s = 0$ state. Microwaves (500 mW) were swept from 2.8 GHz to 2.94 GHz at 1 MHz step. When the microwaves were resonant with the $|0\rangle \leftrightarrow |\pm 1\rangle$ transition, corresponding to 2.87 GHz, the darker $|\pm 1\rangle$ states populated and a drop in the FL was recorded. We were particularly interested in measuring the reduction in the contrast with high amounts of paramagnets. This phenomenon is indicative of a low polarization of the $m_s = 0$ level or, equivalently, a fast relaxation of the $m_s = 0$ level induced by the magnetic noise coming from the spin bath.

**Experimental acquisition time**

In the FL decay experiments, we sampled the dark time interval using $N_p$ exponentially-spaced points. The initialization time $\tau_{in}$ and the readout time $\tau_{ro}$ were kept constant for all the sampling points. Then, the duty cycle time can be calculated:

$$T_{dc} = N_p(\tau_{in} + \tau_{ro}) + T_{min} \frac{\left(\frac{T_{max}}{T_{min}}\right)^{\frac{N_p}{N_p-1}} - 1}{\left(\frac{T_{max}}{T_{min}}\right)^{\frac{1}{N_p-1}} - 1} \qquad (4)$$

where $T_{min}$ and $T_{max}$ are respectively the minimum and the maximum values of the dark time interval. The duty cycle was repeated from several hundred to several thousands of times to increase the SNR ratio. Typical acquisition parameters were $T_{min} = 1$ μs, $T_{max} = 20$ ms, $N_p = 20$, $\tau_{in} = 500$ μs and $\tau_{ro} = 5$ μs, giving $T_{dc} \approx 49$ ms. As an example, a thousand iterations of the duty cycle result in a total acquisition time of 49 s for a full curve.

**Preparation of the NDs mixture with gadoteridol**

The two unfunctionalized NDs were supplied in water, at a concentration of 1 mg/mL. The silica-coated NDs were supplied in ethanol, at the same concentration. Suspensions of NDs were hand shaken and then sonicated for several minutes to disperse NDs. In the first experimental session, we prepared several solutions of $Gd^{3+}$ complexes at different concentrations by dilution of gadoteridol (Prohance, Bracco Diagnostic Inc, initial concentration 0.5 M) with deionized water. Two μL of each gadoteridol solution were mixed to 4 μL of NDs suspensions (containing approximately $2 \times 10^9$ NDs). The total 6 μL of the mixture were deposited onto glass slides and left to air dry, which results in NDs deposited area of ≈2 mm diameter. The amount of deposited gadoteridol varied from 1 pmol to 100 nmol. For each sample fluorescence was recorded on several $(10 - 20)$ different spots. The 40 nm NDs underwent a similar mixture and deposition, while 2 μL of the silica-coated NDs were firstly taken from the ethanol bath and mixed with 2 μL of deionized water before the same kind of processing. The distribution of NDs on the glass slide was checked with a wide-field fluorescence microscope (Nikon Eclipse Ti-E).

**Preparation of the NDs mixture with magnetosomes (MNs)**

We mixed NDs with iron oxide magnetic nanoparticles extracted from magnetotactic bacteria, as described in Mannucci et. al.[30]. MNs suspended in saline buffer were precipitated and concentrated by exposing the vial to a magnetic field generated by a small permanent magnet. Different volumes (1, 2 and 4 μL) of highly concentrated nanoparticles were sampled and mixed with a 2 μL sample of 100 nm uncoated NDs. The new mixture was then deposited on a circular glass slide and allowed to dry before acquiring the FL.

Part of the MNs suspension was mixed with an acid and spectroscopic analyses determined the concentration of dissolved iron. Since all the iron was bound to oxygen in the magnetite phase ($Fe_3O_4$), from the iron (Fe) concentration it was possible to extract the number of MNs, considering a standard MN size of 40 nm.

**Preparation of the NDs mixture with deoxygenated blood and heparin**

Freshly drawn deoxygenated rat blood was collected in heparinized centrifuge tubes to avoid sample coagulation. Two µL of blood were then mixed with either 2 µL of saline solution or 2 µL of deionized water. The mixture with water was intended to induce hemolysis and release paramagnetic deoxyhemoglobin. Then we added 2 µL of uncoated NDs to each solution. The two liquid volumes were deposited onto glass slides, then dried, and finally the FL profile was recorded.

## Author Contributions

# Fast and sensitive detection of hemoglobin and other paramagnetic species using coupled charge and spin dynamics in strongly fluorescent nanodiamonds

## Supplementary Material

F. Gorrini, R. Giri, C. E. Avalos, S. Tambalo, S. Mannucci, L. Basso, N. Bazzanella, C. Dorigoni, M. Cazzanelli, P. Marzola, A. Miotello and A. Bifone

**Characteristics of the three NDs types and estimate of dephasing time**

The three types of NDs are indicated in Table S1. The reported average diameters an estimate, as the NDs have irregular shapes. Diameters measured by scanning electron microscopy (SEM) and dynamic light scattering (DLS) were provided by Bikanta. SEM experiments were also performed in house for further characterization. $T_2^*$ was estimated from the full-width-half-maximum (FWHM) of optically detected magnetic resonance (ODMR) and from the free induction decay (FID). In the first procedure, the inhomogeneous broadening $\gamma_{in}$ of ODMR signal was measured at various microwaves power intensities. Then $T_2^*$ was calculated according to the relation $T_2^* = 1/(\pi \gamma_{in}^0)$, where $\gamma_{in}^0$ is the extrapolated $\gamma_{in}$ at zero microwaves field. The second procedure relied on measuring FID of a spin ensemble through Ramsey interferometry. The system was initially prepared in the state $|0\rangle$ and then flipped into the superposition $\frac{1}{\sqrt{2}}(|0\rangle + |-1\rangle)$ with a $\frac{\pi}{2}$ microwave pulse. The FID of the spin ensemble was measured optically. We fitted the decay profile with a single exponential to extract $T_2^*$. The two estimates are in good agreement and return a short value of $\approx 70$ ns, as expected for high-density NV centers.

| ND Type | Diameter (SEM) [nm] | Diameter (DLS) [nm] | Coating | $T_2^*$ (ODMR FWHM) [ns] | $T_2^*$ (FID) [ns] |
|---|---|---|---|---|---|
| 40 nm bare | 42 ± 16 | 107 ± 0.8 | None | 50 ± 1 | -- |
| 100 nm bare | 120 ± 37 | 188 ± 7 | None | 70 ± 2 | -- |
| 100 nm SiO$_2$ | 120 ± 37 core 9 ± 2 shell | -- | Silica | 70 ± 1 | 78 ± 6 |

**Table S1 |** Size of the NDs, measured by SEM and DLS, functionalization of the surface and $T_2^*$ estimated from the inhomogeneous broadening of the ODMR and from Ramsey interferometry.

**Repolarization dynamics**

In this paragraph we want to show that lasers pulses up to 50 μs do not repolarize the ensemble of NV centers, coherently with previous observations[1]. To prove that, we employed the pulse sequence of Fig. S1. An initializing 532 nm laser pulse of 500 μs and 2 mW of power polarizes the NV ensemble into the $|0\rangle$ state. The $|-1\rangle$ state was optionally populated with the application of a microwave $\pi$ pulse. We determined $\pi = 160$ ns from Rabi oscillations between the states $|0\rangle$ and $|-1\rangle$, at earth magnetic field. The photon counter gate was opened for 300 ns after a variable $\tau_l$ time from the switching on of the readout pulse. Fig. S1 shows the repolarization effect due to the readout pulse on the $|0\rangle$ and $|-1\rangle$ states (black and red curve, respectively), after $\tau_l$. One can see that repolarization does not occur even after tens of microseconds. The FL contrast is proportional to the difference of the curves and remains high even after 50 μs. This justifies the choice of 1 μs of readout time in most of the experiments and 50 μs in the ultrafast single-point acquisition scheme.

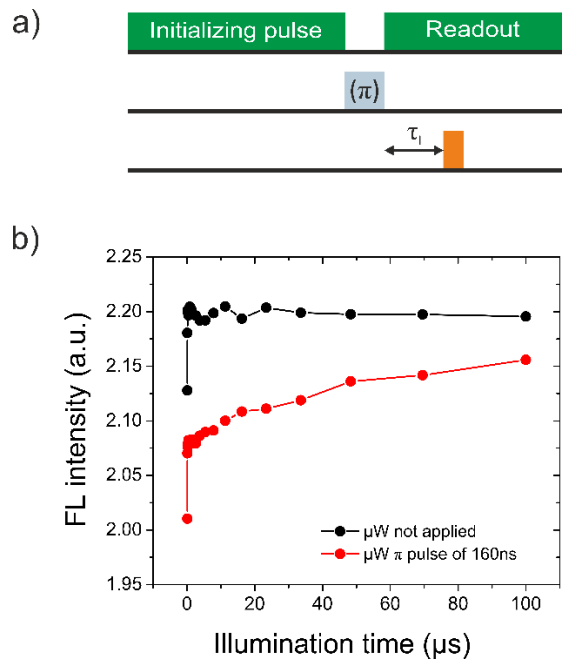

**Figure S1 | Spin repolarization induced by the readout pulse.** (a) Sequence used to determine the repolarization time. After the initialization pulse and an optional microwave $\pi$ pulse, the gate is opened for 300 ns, after a variable $\tau_l$ time from the switching on of the readout pulse. (b) Effect of the readout pulse on the system initially in the $|0\rangle$ state (black curve) and in the $|-1\rangle$ state (red curve). In the whole time interval under consideration the readout pulse does not repolarize considerably the system.

**Population after the initialization pulse**

We can describe the relaxations of an ensemble of NV centers, all in the same conditions, with a four-level scheme with transition rates as indicated in Fig. S2. The three lowest levels represent the states $m_s = -1, 0, +1$ of the ground state. The fourth "metastable" level represents the excited state, intermediate singlet states and defects state that can temporarily capture photoexcited electrons from NV⁻. The normalized populations of each level are thus indicated with $n_0$, $n_{-1}$, $n_{+1}$ and $n_m$. The laser excites the system at a rate proportional to its intensity, $I\sigma$. From the metastable state, the system relaxes to the ground state with rates $k$ (to $m_s = 0$) and $k'$ (to $m_s = \pm 1$). Since the metastable level contains excited and intermediate singlet states, $k$ and $k'$ should describe radiative as well as non-radiative decays, such as the spin-dependent coupling with the singlet state (a mechanism known as intersystem crossing, ISC). For this reason, $k \neq k'$. After the initialization pulse, the population of the three levels with $m_s = -1, 0, +1$ tend to their equilibrium value at a rate rate $\gamma$. Solving the rate equation system gives three exponentials with characteristic times $T_1 = (3\gamma)^{-1}$, $T_r = (k + 2k')^{-1}$ and $T_{mix}^{\pm 1} = (\gamma)^{-1}$. $T_1$ is the longitudinal relaxation time, $T_r$ is the recharge time and $T_{mix}^{\pm 1}$ governs a net transfer of population between $m_s = \pm 1$. We assume that $n_{-1} = n_{+1}$, and $T_{mix}^{\pm 1}$ does not play a role. Populations evolve in time according to the relations:

$$\begin{cases} n_0(t) = \frac{1}{3} + \left(n_0(0) - \frac{1}{3} + \frac{k-\gamma}{k+2k'-3\gamma}n_m(0)\right)e^{-\frac{t}{T_1}} - \frac{k-\gamma}{k+2k'-3\gamma}n_m(0)e^{-\frac{t}{T_r}} \\ n_{\pm 1}(t) = \frac{1}{3} - \frac{1}{2}\left(n_0(0) - \frac{1}{3} + \frac{k-\gamma}{k+2k'-3\gamma}n_m(0)\right)e^{-\frac{t}{T_1}} - \frac{k'-\gamma}{k+2k'-3\gamma}n_m(0)e^{-\frac{t}{T_r}} \quad (S1)\\ n_m(t) = n_m(0)e^{-\frac{t}{T_r}} \end{cases}$$

The total fluorescence (FL) intensity, however, depends not only on the populations, but also on the different probability of undergoing radiative/non-radiative transitions for $m_s = 0$ and $m_s = \pm 1$. Because of ISC, the FL intensity of the level $m_s = 0$, $l_o$ is larger than $l_1$, the FL of the levels $m_s = \pm 1$. We set $l_1 = q l_o$ and $k' = qk$, with $q \approx 0.7$, similarly to Ref[2]. The overall FL intensity is then

$$I(t) = I_0\left(1 + \beta e^{-\frac{t}{T_1}} - \alpha e^{-\frac{t}{T_r}}\right) \quad (S2)$$

with two pre-exponential coefficients related to initial populations according to the relations

$$\begin{cases} \alpha = \dfrac{2q(1+2q)T_r - 3(1+2q^2)T_1}{(1+2q)^2(T_r - T_1)} n_m(0) \\ \beta = \dfrac{3(1-q)}{1+2q}\left(n_0(0) - \dfrac{1}{3} + \dfrac{(1+2q)T_r - 3T_1}{3(1+2q)(T_r - T_1)} n_m(0)\right) \end{cases} \quad (S3)$$

Let's notice that equation (S2) describes the FL profile of an ensemble of NV centers all in the same condition, i.e. with identical $k$, $k'$ and $\gamma$ for all the NVs. We will see in the next paragraph that the magnetic noise experienced by the NVs depend on their distance from the NDs surface. To accommodate the resulting distribution in $T_1$, we introduced stretched exponentials (equation (1) of the main article).

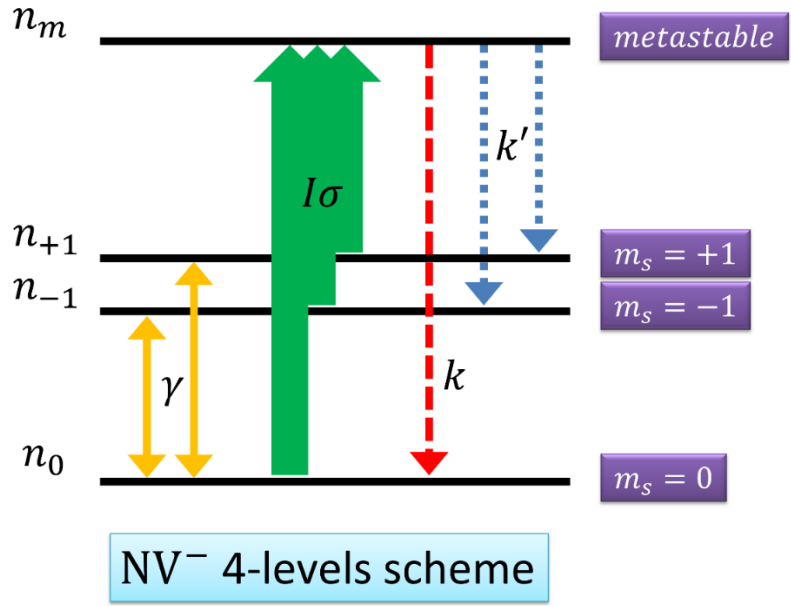

**Figure S2 | 4-level scheme describing the mechanisms of polarization and ionization.** At equilibrium, with the laser off, the population of an NV ensemble is equally distributed between the $|0\rangle$ and $|\pm 1\rangle$ states. When laser is on, the population is pumped into a "metastable" state $|m\rangle$ at a rate $I\sigma$. $|m\rangle$ represents a collection of states (excited states, conduction band, singlet dark states), so $I\sigma$ accounts for both excitation and ionization. From $|m\rangle$ the system can relax to the $|0\rangle$ and $|\pm 1\rangle$ states with decay rates $k$ and $k'$, respectively. Populations in the ground state triplet relax by direct transitions between the $|0\rangle$ and $|\pm 1\rangle$ states (at a rate $\gamma$).

Two things should be remarked about equations (S3). First, both $\alpha$ and $\beta$ depend on $n_{Gd}$. In fact, even if $T_r$ can be fixed to a common value of ≈3 ms, $T_1$ does depend on the amount of Gd and, in turn, also $\alpha$ and $\beta$. This observation justifies treating $\alpha$ as a free parameter when fitting experimental curves, differently from $T_r$ and the stretching factor $m$, that are assumed constant. Second, it is possible to extract the population of the levels right after the laser initialization, at $t = 0$, from $\alpha$ and $\beta$. Reversing equation (S3) we obtain:

$$\begin{cases} n_m(0) = \dfrac{(1+2q)^2(T_r - T_1)}{2q(1+2q)T_r - 3(1+2q^2)T_1} \alpha \\ n_0(0) = \dfrac{1+2q}{3(1-q)}\beta + \dfrac{1}{3} - \dfrac{(1+2q)T_r - 3T_1}{3(1+2q)(T_r - T_1)} n_m(0) \\ n_{\pm 1}(0) = \dfrac{1 - n_0(0) - n_m(0)}{2} \end{cases} \quad (S4)$$

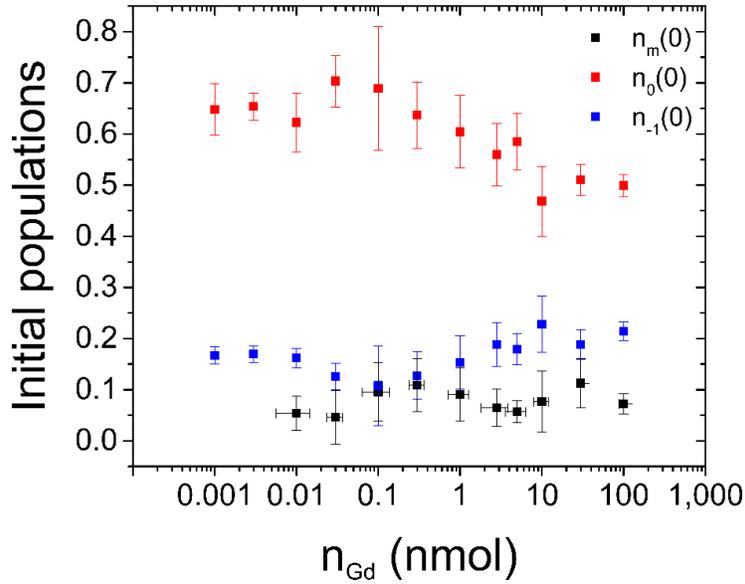

**Figure S3 | Initial populations, after the initialization pulse.** $n_0$ indicates the polarized fraction of the NVs, while $n_{-1} = n_{+1}$ represent the population in the $|\pm 1\rangle$ states of the ground state triplet. $n_m$ indicates the fraction of NVs that have been ionized. We see that $n_m$ does not change substantially with the amount of deposited Gd, while the polarization decreases at high $n_{Gd}$. This is consistent with the observed reduction in the ODMR contrast.

Populations as a function of $n_{Gd}$ are plotted in Fig. S3. Below 10 pmol, $n_0$ and $n_\pm$ saturate, while $n_m$ cannot be determined because the recharge component in the FL spectrum is completely overcome by spin relaxation. At higher concentrations, the degree of polarization decreases, and the difference between $n_0$ and $n_{\pm 1}$ is reduced. This is in good qualitative agreement with the observed reduction in the ODMR contrast. $n_m$ does not change in the considered range of $n_{Gd}$. Indeed, this suggest that ionization is independent of the amount of Gd spins in the surroundings.

**Computation of $T_1$ of NV ensemble in presence of gadoteridol**

The additional relaxation term in equation (1) of the main article contains two quantities that depend on the amount of gadoteridol: the variance of the transverse magnetic field $\langle B_\perp^2 \rangle$ and the rate of Gd fluctuations $f_{Gd}$. The magnetic dipolar field generated by a single Gd ion at a position $\mathbf{R}'_i$ from the NV center is

$$\mathbf{B}_i = \frac{\mu_0 \gamma_e \hbar}{4\pi R'^3_i} \left[ \mathbf{S}_i - \frac{3(\mathbf{S}_i \cdot \mathbf{R}'_i)\mathbf{R}'_i}{R'^2_i} \right] \tag{S5}$$

so the variance $\langle B_{\perp,i}^2 \rangle$ is $\langle B_i^2 - B_{z,i}^2 \rangle$. The following theoretical model is derived as in Ref[2]. We take the average of $\mathbf{B}_i$ over the possible values of $\mathbf{S}_i$ by taking the trace with a purely mixed state, with density matrix $\rho = \frac{1}{2S+1} \mathbb{1}_{2S+1}$, where $\mathbb{1}_{2S+1}$ is the identity matrix of size $2S+1$. Then

$$\langle B_{\perp,i}^2 \rangle = \mathrm{Tr}\{\rho(B_i^2 - B_{z,i}^2)\} = \left(\frac{\mu_0 \gamma_e \hbar}{4\pi}\right)^2 S(S+1) \frac{5 - 3\cos^2 \Theta'}{3 R'^6_i} \tag{S6}$$

where $\Theta'$ is the angle between $\mathbf{R}'_i$ and the quantization axis of the NV center, taken along the z-axis. The origin of the coordinate axis is at the center of the ND. A generic NV center has spherical coordinates $(r, \vartheta, \varphi)$ and a generic Gd ion has spherical coordinates $(R, \Theta, \Phi)$. The noise variance $\langle B_{\perp,i}^2 \rangle$ depends also on the coordinates of the NV center inside the ND, through the relations $R'^2_i = (\mathbf{R} - \mathbf{r})^2$ and $R' \cos \Theta' = R \cos \Theta - r \cos \vartheta$. The total magnetic field on a particular NV center takes the contribution of all the spins surrounding each ND, with a density of spin per unit volume $n_{Gd}$. This means to integrate eq S6:

$$\langle B_\perp^2 \rangle = \sum_i \langle B_{\perp,i}^2 \rangle \approx \left(\frac{\mu_0 \gamma_e \hbar}{4\pi}\right)^2 S(S+1) n_{Gd} \int_{R_{ND}}^{+\infty} dR\, R^2 \int_0^\pi d\Theta \sin\Theta \int_0^{2\pi} d\Phi \frac{5 - 3\cos^2 \Theta'}{3 R'^6} \tag{S7}$$

The result is independent on the azimuthal angle $\varphi$ and depends only on the polar angle $\vartheta$ and on the reduced radius $\xi = r/R_{ND}$:

$$\langle B_\perp^2 \rangle(\xi, \vartheta) = C_1 \left[ \frac{3 - \cos^2 \vartheta}{(1 - \xi^2)^3} + \frac{1 - 3\cos^2 \vartheta}{8\xi^2} \left( \mathrm{atanh}(\xi) - \frac{\xi(1 + \xi^2)}{(1 - \xi^2)^2} \right) \right] \tag{S8}$$

where $C_1 = \frac{2\pi}{3R_{ND}^3}\left(\frac{\mu_0 \gamma_e \hbar}{4\pi}\right)^2 S(S+1)n_{Gd}$. The value of $\xi$ runs up to a maximum value of $\approx 0.95$, so it assumes that NV centers too close to the surface (within 2-3 nm) are not stable in the negative form.

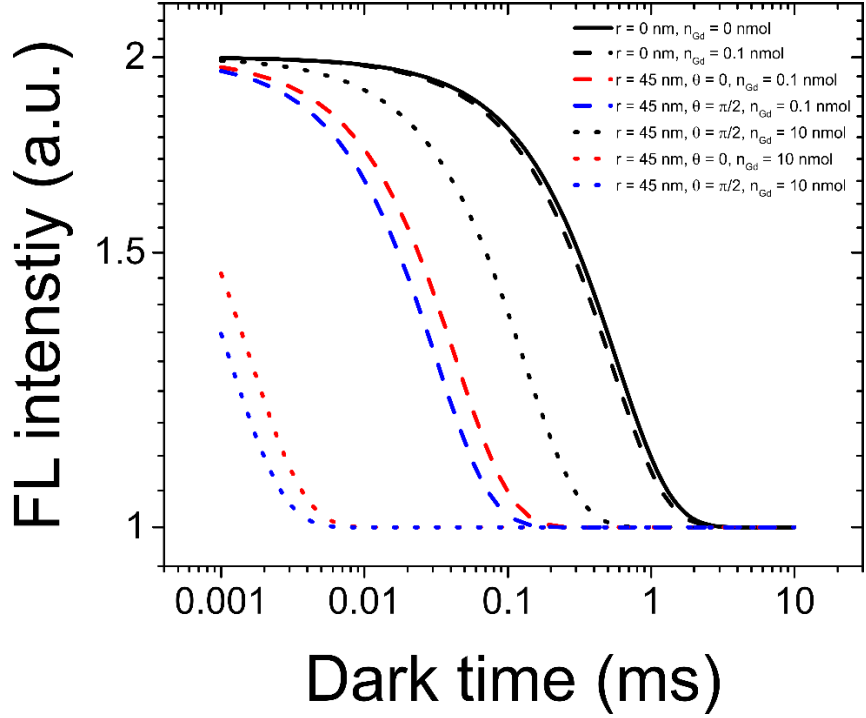

**Figure S4 | Simulation of the spin-related decay of FL for a single NV center inside a ND.** The symmetry axis of the NV is along the z-axis. Black curves: FL decay profile for an NV at the center of the ND, with no Gd (solid line), 0.1 nmol (dashed line), and 10 nmol (dotted line) of Gd deposited. Red curves: luminescence decay for an NV at 45 nm from the center, along the z axis ($\vartheta = 0$), with 0.1 nmol and 10 nmol of Gd (dashed and dotted lines, respectively). Blue curve: same luminescence decay profile for an NV at 45 nm from the center, on the equatorial plane ($\vartheta = 90°$), with 0.1 nmol and 10 nmol of Gd (dashed and dotted lines, respectively). The red and blue solid lines coincide with the black one: without Gd, all the NV centers have the same $T_1$, since no relaxation mechanism other than Gd-paramagnetic noise has been considered here (no cross-relaxation and surface charges effect).

The rate of Gd fluctuations $f_{Gd}$ includes a constant vibrational term[3] $f_{vib} \approx 1$ GHz and a dipolar term $f_{dip}$ that depends on the amount of Gd ions. The dipolar interaction between a Gd spin $\mathbf{S}_i$ with all the other spins $\mathbf{S}_j$ of the bath gives an estimate of $f_{dip}$. Explicitly: $hf_{dip} = \sqrt{\sum_{j \neq i}\langle H_{ij}^2\rangle}$, where $H_{ij}$ is the interaction between two magnetic dipoles

$$H_{ij} = \frac{\mu_0 \gamma_e^2 \hbar^2}{4\pi R_{ij}^3} \left[ \mathbf{S}_i \cdot \mathbf{S}_j - \frac{3(\mathbf{S}_i \cdot \mathbf{R}_{ij})(\mathbf{S}_j \cdot \mathbf{R}_{ij})}{R_{ij}^2} \right] \qquad (S9)$$

where $\mathbf{R}_{ij}$ is the vector connecting the two spins. The quantity $\langle H_{ij}^2 \rangle$ is equal to $\text{Tr}\{\rho H_{ij}^2\}$, with the two-spins density matrix $\rho = \frac{1}{(2S+1)^2} \mathbb{1}_{2S+1} \otimes \mathbb{1}_{2S+1}$. So

$$\langle H_{ij}^2 \rangle = \left( \frac{\mu_0 \gamma_e^2 \hbar^2}{4\pi} \right)^2 \frac{2 S^2 (S+1)^2}{3 R_{ij}^6} \qquad (S10)$$

and once again, the summation over j can be replaced by an integral:

$$\sum_{j \neq i} \langle H_{ij}^2 \rangle \approx \left( \frac{\mu_0 \gamma_e^2 \hbar^2}{4\pi} \right)^2 \frac{2 S^2 (S+1)^2}{3} n_{Gd} \int_{R_{min}}^{R_{max}} dR_{ij} 4\pi R_{ij}^2 \frac{1}{R_{ij}^6} \qquad (S11)$$

If $R_{min}$ is about the size of a gadoteridol molecule, $d_{Gd}$, and $R_{max} \gg R_{min}$, then

$$f_{dip} \approx \frac{\mu_0 \gamma_e^2 \hbar}{6\sqrt{2\pi^3}} S(S+1) \sqrt{\frac{n_{Gd}}{d_{Gd}^3}} \qquad (S12)$$

Finally, the reduction of $T_1$ due to Gd spins can be described by

$$\frac{1}{T_1}(\xi, \vartheta) = \frac{1}{T_1^0} + k_1(\xi, \vartheta) \frac{n_{Gd}(\sqrt{n_{Gd}} + k_2)}{n_{Gd} + 2k_2 \sqrt{n_{Gd}} + k_3} \qquad (S13)$$

with $k_2 = 0.8423$ nm$^{-3/2}$, $k_3 = 6.5532$ nm$^{-3}$ and

$$k_1(\xi, \vartheta) = 1.1347 \cdot 10^{-5} \text{ GHz nm}^{\frac{3}{2}} \cdot \left[ \frac{3 - \cos^2 \vartheta}{(1 - \xi^2)^3} + \frac{1 - 3\cos^2 \vartheta}{8\xi^2} \left( \text{atanh}(\xi) - \frac{\xi(1 + \xi^2)}{(1 - \xi^2)^2} \right) \right]. (S14)$$

A few examples of FL decay with Gd are shown in Fig. S4.

However, the FL comes from many NV centers in different positions inside each ND. This means integrating numerically the equation (S13) over $r$, $\vartheta$ and $\beta$:

$$I(t, n_{Gd}) = \int_0^{r_{max}} dr\, r^2 \int_0^\pi d\vartheta \sin\vartheta \int_0^{2\pi} d\varphi\, e^{-\frac{t}{T_1(\xi, \vartheta)}} \sim I_0 e^{-\left(\frac{t}{T_1}\right)^n} \qquad (S15)$$

where $I_0$ is just a coefficient for the luminescence, and the values of the effective longitudinal relaxation time $T_1$ and the stretching coefficient $n$ are obtained by fitting $I(t, n_{Gd})$. Values of $n$ obtained with this procedure are shown in Fig. S5. Even if we do not consider any explicit model for recharge, we expect that the contribution to the growing component of the FL should be described by a stretched exponential, following a similar procedure that led to equation (S15).

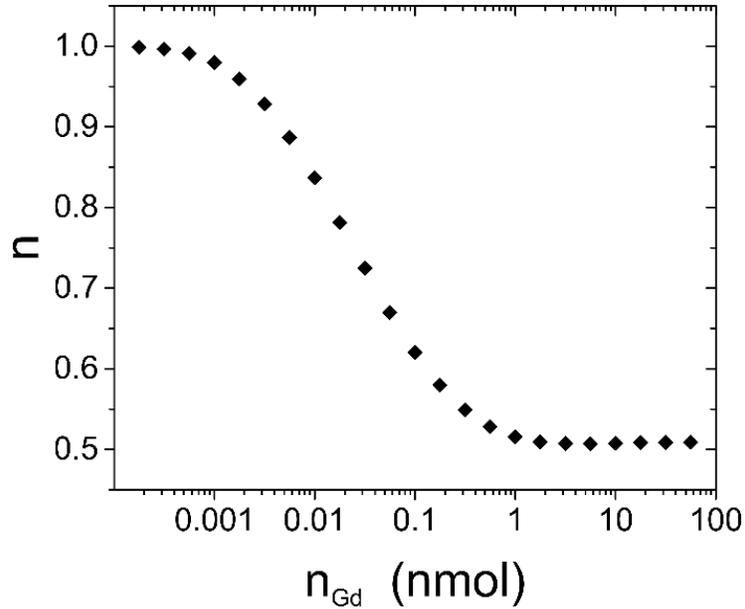

**Figure S5 | Values of $n$ obtained by fitting with equation (S15) the numerically computed FL profile.**

The NDs/Gd mixture is a heterogeneous system, where Gd can both decorate the surface of NDs and also fill the empty spaces between packed NDs. To simplify the derivation of $\langle B_{\perp,i}^2 \rangle$ we considered a single ND, immersed in pure gadoteridol, with no other NDs in the surroundings. Analogously, in the derivation of $f_{dip}$ we considered a homogeneous system of pure gadoteridol, with no NDs at all. This heterogeneity of the system means that the effective amount of Gd interacting with NDs is lower than the one reported. From the best fit of the theoretical curve of Fig. 2a, we empirically found that only $15 - 30\%$ of gadoteridol interacts efficiently with NDs.

**Estimate of sensitivity and acquisition time**

Finally, we estimate the sensitivity of NV ensemble to variations in the amount of deposited gadoteridol. The number of photons detected at a fixed time $t$ after $N_{iter}$ iterations is

$$\mathcal{N}_{ph}(t) \approx \mathcal{R}\tau_{ro}N_{iter}\left(1 + \beta e^{-\left(\frac{t}{T_1}\right)^n} - \alpha e^{-\left(\frac{t}{T_r}\right)^m}\right) \qquad (S16)$$

where $\mathcal{R}$ is the photon counting rate and $\tau_{ro}$ the readout time. The photon shot noise scales as the square root:

$$\delta \mathcal{N}_{noise}(t) \approx \sqrt{\mathcal{R}\tau_{ro}N_{iter}\left(1 + \beta e^{-\left(\frac{t}{T_1}\right)^n} - \alpha e^{-\left(\frac{t}{T_r}\right)^m}\right)} \quad (S17)$$

This noise must be compared with a small variation of the signal due to a change of the relaxation time $\delta T_1$, which, in turn, depends on a variation in the number of external magnetic dipoles:

$$\delta \mathcal{N}_{signal}(t) \approx \mathcal{R}\tau_{ro}N_{iter}\beta n \left(\frac{t}{T_1}\right)^n e^{-\left(\frac{t}{T_1}\right)^n} \frac{\delta T_1}{T_1} \quad (S18)$$

Again, the assumption that recharge is independent of the magnetic noise holds. The signal to noise ratio can be written as:

$$SNR(t) = \frac{\delta \mathcal{N}_{signal}}{\delta \mathcal{N}_{noise}}(t) = \mathcal{C} \frac{\beta n \left(\frac{t}{T_1}\right)^n e^{-\left(\frac{t}{T_1}\right)^n}}{\sqrt{1 + \beta e^{-\left(\frac{t}{T_1}\right)^n} - \alpha e^{-\left(\frac{t}{T_r}\right)^m}}} \frac{\delta T_1}{T_1} \quad (S19)$$

where $\mathcal{C} = \sqrt{\mathcal{R}\tau_{ro}N_{iter}}$ is a parameter that depends only on detection conditions. We see that SNR benefits from the high density of NV centers. In fact, the presence of a recharge mechanism decreases the denominator of equation (S19). Further, the high NV density allows high photon counting rates (up to $\sim 10^7 s^{-1}$) and long readout pulses (up to tens of microseconds), as described above, and consequently leads to a large $\mathcal{C}$ coefficient.

Numerically, we find that the maximum of SNR of equation (S19) is found for $t \approx T_1$, unlike the simple exponential case, where the maximum is at $\frac{T_1}{2}$. Explicitly,

$$SNR_{max} \approx \mathcal{C} \frac{\beta n e^{-1}}{\sqrt{1 + \beta e^{-1} - \alpha e^{-g}}} \frac{\delta T_1}{T_1} \quad (S20)$$

where $g = \left(\frac{T_1}{T_r}\right)^m$. In a single-point detection scheme, $SNR_{max}$ is obtained after a total experimental time of

$$\Delta t_{exp} = N_{iter}(\tau_{in} + \tau_{ro} + T_1) = \frac{\mathcal{C}^2}{\mathcal{R}\tau_{ro}}(\tau_{in} + \tau_{ro} + T_1) \quad (S21)$$

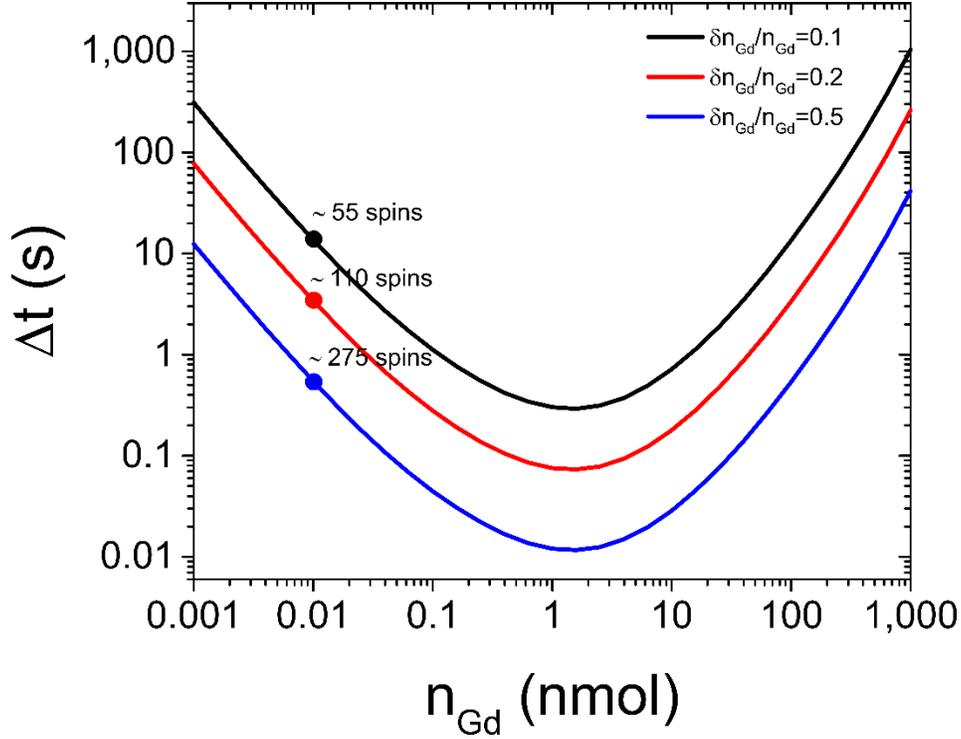

**Figure S6 | Acquisition time needed to sense variation of 10% (black curve), 20% (red curve) and 50% (blue curve) of deposited gadoteridol.** At 10 pmol, ≈0.5 s can detect an increment of 275 Gd spins over 550 total spin per each ND. Subnanomolar amounts can be detected in few tens of milliseconds, around the minimum of the curves.

Choosing $SNR_{max} = 1$, it is possible to relate $\Delta t_{exp}$ to a variation $\delta T_1$ of the relaxation time, or, equivalently, to a relative change in the number of magnetic dipoles, expressed as

$$\frac{\delta c_{Gd}}{c_{Gd}} = \frac{dc_{Gd}}{dT_1}\frac{\delta T_1}{c_{Gd}} \qquad (S22)$$

where now $c_{Gd}$ is considered a monotonic function of $T_1$ (Fig. 2a of the main article). Finally

$$\Delta t_{exp} = \frac{\mathcal{K}_{det}\mathcal{K}_{sample}}{\left(\frac{\delta c_{Gd}}{c_{Gd}}\right)^2} \qquad (S23)$$

with a detector-related coefficient (in units of s)

$$\mathcal{K}_{det} = \frac{\tau_{in} + \tau_{ro} + T_1}{\mathcal{R}\tau_{ro}} \qquad (S24)$$

and a sample-dependent coefficient

$$\mathcal{K}_{sample} = \left(\frac{dc_{Gd}}{dT_1}\frac{T_1}{c_{Gd}}\right)^2 \frac{e(e+\beta-\alpha e^{1-g})}{(\beta n)^2} \tag{S25}$$

In Fig. S6 the total experimental time is plotted as a function of a relative variation of 10%, 20% and 50% at various concentrations of Gd, assuming $\mathcal{R} = 10^7 \text{s}^{-1}$ and $\tau_{ro} = 50$ µs. Subnanomolar concentrations can be detected in $\approx 10$ ms (minimum of the blue curve). Let's notice that the steepest slope of $T_1$ (or equivalently, of $t_{min}$ and $\beta - \alpha$) coincides with the range of the shortest acquisition time.